\documentclass{ifacconf}

\usepackage{graphicx}      
\usepackage{natbib}        
\usepackage{amsmath}
\usepackage{amssymb}
\usepackage{algorithmic}
\usepackage{textcomp}
\usepackage{color}
\usepackage{tikz}
\usepackage{mathtools}
\usepackage{pgfplots}
\usepackage{siunitx}
\usepackage{lipsum}
\usetikzlibrary{shapes,arrows}
\usepackage{subcaption}

\begin{document}
\begin{frontmatter}

\title{The impact of load placement on grid resonances during grid restoration} 

\author[First]{Adolfo Anta}
\author[First]{Diego Cifelli}
\address[First]{AIT Austrian Institute of Technology GmbH, Vienna, Austria (e-mail: \{adolfo.anta,diego.cifelli\}@ait.ac.at)}

\begin{abstract}
As inverter-based generation is being massively deployed in the grid, these type of units have to take over the current roles of conventional generation, including the capability of restoring the grid. In this context, the resonances of the grid during the first steps of a black start can be concerning, given that the grid is lightly loaded. Especially relevant are the low frequency resonances, that may be excited by the harmonic components of the inverter. A typical strategy to avoid or minimize the effect of such resonances relies on connecting load banks. This was fairly feasible with conventional generation, but given the limited ratings of inverters, the amount of load that can be connected at the beginning is very limited. In this paper we consider the energization of a transmission line, and investigate the optimal location of a load along a line in order to maximize the damping in the system. By analysing the spectral properties as a function of the load location, we formally prove that placing the load in the middle of the transmission line maximizes the damping ratio of the first resonance of the system. 


\end{abstract}


\end{frontmatter}

\section{Introduction and motivation}
Traditionally, grid restoration has relied on black-start capable power plants. As inverters are populating the grid and conventional generation is being decommissioned, the classical roles and functionality have to be transferred, including the ability to restart a grid~(\cite{noris2019power, ng_restart}). Even in the cases where grid restoration plans rely mainly on hydro power plants, operators may still want to replace them by inverters to speed up the process~(\cite{aniceto2023towards}). This also presents an opportunity to upgrade grid restoration techniques, given the controllability and flexibility of inverter-based generation. Strategies used to rely on a divide-and-conquer approach, where the grid is split in several regions or cells, that are independently reenergized and synchronized afterwards. However, a faster and simpler strategy would define a single backbone covering a large part of the grid that is energized at once. One of the drawbacks of the backbone-based solution is the emergence of resonances, given that during the first steps the grid is barely loaded. Resonances are more likely to appear in larger grids, leading to overvoltages as they are excited by the harmonic components of the inverter.

Among other strategies, nowadays it is common for operators to slow down energization ramps to avoid triggering resonances. Although this may work in some cases, it slows down the process and may cause some issues with protection devices. Other possible solution relies on adjusting the voltage gains in the grid-forming energizing inverters. This would lead to poor tracking of the voltage reference (typically a step or a ramp), which may jeopardize the grid restoration process. Another strategy to improve the dynamic characteristics of the system during grid restoration and avoid resonances relies on connecting load banks, which clearly improves the overall damping in the system. Given the large ratings of conventional power plants, finding out which load to connect and where to was not so critical. Moreover, resonances were not so relevant since conventional generation would not excite high frequencies. Hence, the limited ratings of inverter-based generation makes the optimal location of load banks a very relevant question. The load value cannot be freely chosen, and in any case it is limited because of the limited ratings of the inverters, and the (mainly reactive) power demand of the backbone to be energized. Moreover, as we will see, just placing a large load does not remove all resonances, regardless of its value. Although the exact location of the load to be connected may not be completely chosen, it would be very beneficial to gain intuition on which locations are preferable, thereby creating simple, intuitive rules for the system operator.


There is relatively little work in the field of optimal placement of loads. The article  in~\cite{hiskens1997locating}, while not focused on black start, proposes a numerical method to determine which load plays a significant role, by means of computing the sensitivities. The damping of a grid has been studied in~\cite{mallada2011improving} and~\cite{borsche2015effects}, but focusing on conventional generation and its corresponding swing equation model, while skipping line dynamics. Resonances are also widely studied in the context of converters connected to a grid via long transmission lines, but to design adequate converter controls to damp resonances (e.g., ~\cite{zhang2013resonance}). 

In this article we consider a backbone consisting of a long transmission line, and derive analytically the spectra of the system as a function of the load location. While this is a relatively simple setup, certain system operators are already energizing long lines using storage units and facing resonances. Leveraging existing results for 2-Toeplitz matrices, analytical expressions for the characteristic polynomials defining the eigenvalues are derived. We formally prove how the optimal location of the load is precisely the middle point of the transmission line to be energized, \textit{if} the goal is to maximize the damping of the resonant mode corresponding with the lowest frequency. An optimal location that maximizes the damping for all resonances does not exist, and thus the solution depends on the resonances of interest. Likewise, pertinent observations about the evolution of resonances as a function of the load value are drawn. It is also pointed out that a large load does not manage to avoid all resonances, and, in fact, certain resonances are not controllable depending on the location of the load. It can be concluded that having loads distributed along the line is more effective at reducing the resonances in the grid, rather than placing a large load at a single point.

\begin{figure*}[h]
\includegraphics[width=\linewidth]{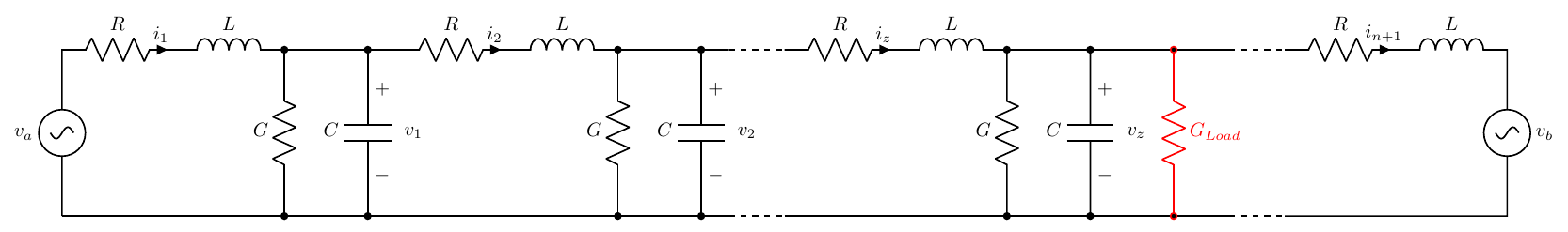}
\caption{Transmission line model with cascaded $\pi$ networks and a load connected at section $z$.}
\label{fig:PiModel}
\end{figure*}

\section{Modelling}
We consider the energization of a long transmission line (see Figure~\ref{fig:PiModel}) representing the grid backbone. The line model corresponds to a cascaded connection of $\pi$-sections, that allow us to represent the dynamics through a linear state space model and a finite number of states~(\cite{macias2005comparison}). For simplicity, a single phase system is selected, although using modal decomposition the same concepts can be applied for 3-phase systems. At both ends of the line there could be voltage sources, representing grid-forming devices, being responsible for the energization of the line. The diagram depicts as well an added load with conductance $G_{load}$, located along the line at the section $z$. Writing one differential equation per dynamic element (capacitors and inductances) leads us to a standard linear representation of the type $\dot x = A x +B u$ such as in~\eqref{eq:modelPi_withLoad}, with $i$ and $v$ denoting the current and the voltage for each branch and node, $v_a$, $v_b$ the voltage sources, and $R$, $G$, $L$ and $C$ representing the resistance, conductance, inductance and capacitance per $\pi$-section. At the element $(2z,2z)$ of the $A$ matrix we see the effect of the added load, with $G_L=G+G_{Load}$. Non-mentioned entries in the $A$ matrix in~\eqref{eq:modelPi_withLoad} are equal to zero. The dimensions of the $A$ matrix are $2n+1$, where $n$ is the total number of lumped sections.

\begin{figure*}[h]
\begin{align}
\frac{d}{dt} \left[ \begin{array}{c}
i_1  \\ v_1 \\ i_2 \\ \vdots \\ i_{z-1} \\ v_z \\ i_z \\ \vdots \\ v_n \\ i_{n+1}   
\end{array}\right]=&
\left[\begin{array}{rrrrrrrrrrr}
-R/L & -1/L &        &        &        &        & &&&\\
1/C  & -G/C & -1/C   &        &        &        & &&&\\
     &  1/L & -R/L   & -1/L   &        &        &        &        &        &\\
     &      & \ddots & \ddots & \ddots &        &        &        &        & \\
     &      &        &  1/L   & -R/L   & -1/L   &        &        &        &  \\
     &      &        &        & 1/C    & -G_L/C & -1/C   &        &        &\\
     &      &        &        &        & 1/L    & -R/L   & -1/L   &        &\\    
     &      &        &        &        &        & \ddots & \ddots & \ddots &\\
     &      &        &        &        &        &        & 1/C    &  -G/C  & -1/C\\
     &      &        &        &        &        &        &        &   1/L  &  -R/L
\end{array}\right] 
\left[ \begin{array}{c}
i_1  \\ v_1 \\ i_2 \\ \vdots \\ i_{z-1} \\ v_z \\ i_z \\ \vdots \\ v_n \\ i_{n+1}      
\end{array}\right] +\notag\\
&
\left[\begin{array}{cc}
1/L &  0\\
0  & 0\\
0  & 0\\
\vdots & \vdots\\
0  & 0\\
0 &   -1/L
\end{array}\right] 
\left[ \begin{array}{c} v_a  \\ v_b   
\end{array}\right]
\label{eq:modelPi_withLoad}
\end{align}
\end{figure*}



The $A$ matrix in the state space representation of~\eqref{eq:modelPi_withLoad} is a tridiagonal matrix, and more precisely a 2-Toeplitz matrix. As it will be seen in the next section, the eigenvalues for these matrices can be analytically computed. This type of dynamics has been widely studied, but in general with the focus on analyzing stability properties~(\cite{dorfler2018electrical}) rather than analysing damping levels. It has been conjectured that the damping does depend on the grid topology and the amount of connected loads, but a theoretical analysis is missing. Moreover, how to select the load location to maximize damping ratios is still unknown. 


\section{Spectral analysis and analytical characterization}

To analyze the spectral properties of the system under different load positions, we exploit the similarities between the state space representation in~\eqref{eq:modelPi_withLoad} and Toeplitz matrices. This helps us derive analytical expressions for the system eigenvalues and its dependence on the  placement of loads. 

\subsection{An overview of 2-Toeplitz matrices and Chebyshev polynomials}
We first provide in this subsection a short overview of Toeplitz matrices. We refer the interested reader to~\cite{gover1994eigenproblem, da2007characteristic,da2020comments,alvarez2005some} for more details. A matrix $A$ is tridiagonal if $a_{ij}=0$ whenever $\vert i-j\vert>1$. Moreover, $A$ is r-Toeplitz of order $n$ if $a_{i+r,j+r}=a_{ij}$, for $i,j=1,2,...n-r$. The inverse and spectra of these matrices have been widely studied in mathematics, and has many applications in  topics such as wave dispersion models or fiber optic design~(\cite{al2017pole,bastawrous2022closed}). 

The characteristic polynomial of these matrices can be written as a function of Chebyshev polynomials. We first define the Chebyshev polynomials of second kind, that satisfy the following 3-point recurrent relationship: 
\begin{equation}
    U_{n+1}(\lambda) := \lambda U_{n}(\lambda) -U_{n-1}(\lambda),\qquad \forall n=1,2,\hdots \:\: \lambda \in\mathbb{C}
\end{equation}
with initial conditions $U_0(\lambda)=1$ and $U_1(\lambda)=2\lambda$. Moreover, $U_n$ satisfies the following trigonometric equation:
\begin{equation}
    U_{n}(\lambda) = \frac{\sin ((n+1)\theta)}{\sin (\theta)}, \text{with } \lambda=\cos(\theta), \: 0\leq\theta < \pi \label{eq:trigDefChebyshev}
\end{equation}
For convenience we also define the following functions:
\begin{eqnarray}    
g(\lambda) &:=& \left(\lambda+\frac{R}{L}\right) \left(\lambda+\frac{G}{C}\right)     \label{eq:defG}\\
P_n(\lambda) &:=&\frac{1}{\left(LC\right)^{n}} U_n\left(\frac{1}{2}LC\lambda+1\right) \label{eq:defP}
\end{eqnarray}
Using these definitions, the eigenvalues of a 2-Toeplitz matrix can be written as the solution of the characteristic polynomial
\begin{equation}
    \Delta_{2n+1}(\lambda) := \left(\lambda+\frac{R}{L}\right)P_n(g(\lambda))=0\label{eq:charPol_odd}
\end{equation}
if the dimensions of the matrix are odd, and 
\begin{equation}
    \Delta_{2n}(\lambda) := P_n(g(\lambda))-\frac{1}{LC} P_{n-1}(g(\lambda))=0 \label{eq:charPol_even}
\end{equation}
if the dimensions are even, see~\cite{gover1994eigenproblem,da2007characteristic} for a detailed explanation. Using expressions~\eqref{eq:charPol_odd} and~\eqref{eq:charPol_even}, in the next subsections we derive formulas for the characteristic polynomial of the system in~\eqref{eq:modelPi_withLoad}.

\subsection{The unloaded case}
\label{sec:unloaded}
In the case of the unloaded system ($G_L=G$), the computation of the eigenvalues and its corresponding damping is fairly straightforward. Indeed, it can be seen that the $A$ matrix in~\eqref{eq:modelPi_withLoad} turns exactly into a 2-Toeplitz matrix of odd size, for which the eigenvalues can be directly computed using~\eqref{eq:charPol_odd}. 
 The equation in~\eqref{eq:charPol_odd} lead us to two nonlinear equations, using the definition in~\eqref{eq:trigDefChebyshev} and identifying the argument inside the Chebyshev polynomial $U_n$ in~\eqref{eq:defP} to $\cos(\theta)$:
\begin{eqnarray}
    \Delta_{2n+1}(\lambda)=\left(\lambda+\frac{R}{L}\right)\frac{1}{(LC)^n}\frac{\sin((n+1)\theta)}{\sin(\theta)} = 0\label{eq:1stEqEvals}\\
    \text{with} \qquad \cos(\theta) = \frac{1}{2}LC\left(\lambda+\frac{R}{L}\right)\left(\lambda+\frac{G}{C}\right)+1 \label{eq:2ndEqEvals}    
\end{eqnarray}


Equation~\eqref{eq:2ndEqEvals} is quadratic in $\lambda$, so it can be solved as a function of $\cos(\theta)$. At the same time,~\eqref{eq:1stEqEvals} defines the conditions for $\theta$ plus another solution for $\lambda$, leading to one real eigenvalue and $2n$ complex eigenvalues:
\begin{gather} 
\lambda_0 = \frac{-R}{L},\label{lambdaReal}\\
\lambda_{\pm k} = -\frac{1}{2}\left(\frac{R}{L}+\frac{G}{C}\right)\pm\sqrt{\frac{1}{4}\left(\frac{R}{L}-\frac{G}{C}\right)^2 - \frac{2(1- \cos(\theta_k))}{LC}}\notag\\ \label{lambdaComplex}
\end{gather} 
where $\theta_k=\frac{k \pi}{n+1}, k=1,2\hdots n$. It is clear from this expression that, for the typical values of a transmission line, the imaginary part is much larger than the real part for all the complex eigenvalues, the corresponding damping ratio is very low and therefore resonances appear. Notice that~\eqref{eq:1stEqEvals} is well defined for $\theta=0$, and in particular $\lambda=\lambda_0$ corresponds to $\theta=0$ in~\eqref{eq:2ndEqEvals}.

\subsection{The effect of the load location}
\label{sec:loadPlacement}
We now consider the placement of a load $G_{load}$ at a section $z$. Because of symmetry in the $A$ matrix, we consider only the values $1\leq z\leq \lceil n/2 \rceil$. For ease of notation, we define now the index $j=2z-1$, and thus $G_L$ appears in the element $(j+1,j+1)$ in the $A$ matrix. Expanding along the $j+1$ column in~\eqref{eq:modelPi_withLoad} where $G_L$ is located, the characteristic polynomial in open loop for this system is:
\begin{eqnarray}
    \Delta_{2n+1} &=& \left(\lambda+\frac{G_L}{C}\right)\Delta_j\Delta_{2n-j} + \frac{1}{LC}\Delta_{j-1}\Delta_{2n-j} \notag\\
    &&+ \frac{1}{LC}\Delta_j\Delta_{2n-j-1}    
\end{eqnarray}
Notice that $j$ is always odd, and therefore the expressions of $\Delta_j$, $\Delta_{2n-j}$ are given by~\eqref{eq:charPol_odd}, while the expressions for $\Delta_{j-1}$ and $\Delta_{2n-j-1}$ are defined in~\eqref{eq:charPol_even}. 
For mathematical convenience, and to obtain a single expression independently of the values taken by $j$, we define $P_{-1}=0$ and therefore $\Delta_0 = 1$. 
Using the expressions in~\eqref{eq:charPol_odd} and~\eqref{eq:charPol_even}, the characteristic polynomial of the $A$ matrix can be written as:
\begin{align*}
    \Delta_{2n+1} &= \left(\lambda+\frac{G_L}{C}\right) \left(\lambda+\frac{R}{L}\right) P_{\frac{j-1}{2}} \left(\lambda+\frac{R}{L}\right) P_{\frac{2n-j-1}{2}} \notag \\ 
    &+ \frac{1}{LC} \left(P_{\frac{j-1}{2}} - \frac{1}{LC} P_{\frac{j-3}{2}}\right) \left(\lambda+\frac{R}{L}\right) P_{\frac{2n-j-1}{2}}\notag\\
    &+ \frac{1}{LC} \left(P_{\frac{2n-j-1}{2}} - \frac{1}{LC} P_{\frac{2n-j-3}{2}}\right) \left(\lambda+\frac{R}{L}\right) P_{\frac{j-1}{2}}   \notag\\ 
\end{align*}

It can be easily seen from this expression that, regardless of the location of the load and its value, $\lambda=-R/L$ is still a solution, as in the unloaded case. Extracting the term $\lambda+\frac{R}{L}$ and using the expressions for $P$ in~\eqref{eq:defP}, we can see that the roots of the characteristic polynomial satisfy the following equation\footnote{Notice that $\theta=0$ is a solution of~\eqref{eq:charPol_overJ} but not a root of the characteristic polynomial, by means of L'H\^{o}pital.}:
\begin{align}
    F_1(\theta,G_L,\lambda):&=h(G_L,\lambda) \sin\left(\frac{j+1}{2} \theta\right) \sin\left(\frac{2n-j+1}{2}\theta\right) \notag\\
&-    \sin\left(\frac{j-1}{2}\theta\right) \sin\left(\frac{2n-j+1}{2}\theta\right)\notag\\
&-\sin\left(\frac{j+1}{2}\theta\right) \sin\left(\frac{2n-j-1}{2}\theta\right) = 0 
\label{eq:charPol_overJ}
\end{align}
with $h(G_L,\lambda)=LC\left(\lambda+\frac{G_L}{C}\right) \left(\lambda+\frac{R}{L}\right) +2$. This expression, together with~\eqref{eq:2ndEqEvals}, defines the eigenvalues of the system for all possible locations of the load along the line. 

While a general conclusion on the load location cannot be driven, it will be shown that for realistic values of the load, the first resonance can be damped at most when the load is located exactly in the center. Moreover, the increase of the damping of this mode is monotonic along the line, that is, the load should be placed as close as possible to the middle point of the transmission line.


To derive a local result and analyse the sensitivity of the roots of the characteristic polynomial with respect to the load, we expand each term using a series expansion of the nonlinear expressions in~\eqref{eq:charPol_overJ} and~\eqref{eq:2ndEqEvals}, evaluated around the unloaded case, that is, $x_* =  (\lambda_*,\theta_*, G_L=G)$:
\begin{equation*}
 F_1  \approx F_1(x_*) + \frac{\partial F_1}{\partial \theta}\Bigr|_{\substack{x_*}} \Delta \theta 
    + \frac{\partial F_1}{\partial G_L}\Bigr|_{\substack{x_*}} \Delta G_{L}  + \frac{\partial F_1}{\partial \lambda}\Bigr|_{\substack{x_*}} \Delta \lambda =0
\end{equation*}
Likewise, the expression in~\eqref{eq:2ndEqEvals} relates the angle $\theta$ and the eigenvalues $\lambda$:
\begin{align}
 F_2 &= \cos(\theta) - \frac{1}{2}LC\left(\lambda+\frac{R}{L}\right)\left(\lambda+\frac{G}{C}\right)-1 \notag\\
 & \approx F_2(x_*) + \frac{\partial F_2}{\partial \theta}\Bigr|_{\substack{x_*}} \Delta \theta 
    +    \frac{\partial F_2}{\partial \lambda}\Bigr|_{\substack{x_*}} \Delta \lambda = 0
\end{align}
Similar approaches are common to handle these type of nonlinear expressions that appear in other applications of Toeplitz matrices, such as quantum mechanics~(\cite{ortega2020spectral}). The expressions for $\lambda_*$ and $\theta_*$ are given by~\eqref{lambdaReal} and~\eqref{lambdaComplex}. Solving this set of two equations, we obtain the sensitivity between the change in the eigenvalues and the change in the impedance:
\begin{equation}
\Delta \lambda = \frac{\frac{\partial F_1}{\partial G_L}\Bigr|_{\substack{x_*}}}{\frac{\partial F_1}{\partial \theta}\Bigr|_{\substack{x_*}}\frac{\frac{\partial F_2}{\partial \lambda}\Bigr|_{\substack{x_*}}}{\frac{\partial F_2}{\partial \theta}\Bigr|_{\substack{x_*}}}-\frac{\partial F_1}{\partial \lambda}\Bigr|_{\substack{x_*}}}\Delta G_L
\label{eq:sensitivities}
\end{equation}
The partial derivatives of the functions $F_1$ and $F_2$ are:
\begin{align}
\frac{\partial F_1}{\partial G_L} &= L\left(\lambda+\frac{R}{L}\right)\frac{1}{2}\left(\cos((n-j)\theta)-\cos((n+1)\theta)\right) \notag\\
\frac{\partial F_1}{\partial \lambda} &= LC\left(2\lambda+\left(\frac{G_L}{C}+\frac{R}{L}\right)\right)\frac{1}{2}(\cos((n-j)\theta)\notag\\
&-\cos((n+1)\theta)) \notag\\
\frac{\partial F_1}{\partial \theta} &= \left(LC\left(\lambda+\frac{G_L}{C}\right) \left(\lambda+\frac{R}{L}\right) +2\right)\frac{1}{2}\notag\\
&\left(-(n-j)\sin((n-j)\theta)+(n+1)\sin((n+1)\theta)\right)\notag\\
&-n\sin(n\theta)-\frac{1}{2}(-(n-j+1)\sin((n-j+1)\theta)\notag\\
&-(n-j-1)\sin((n-j-1)\theta))\notag\\
\frac{\partial F_2}{\partial \theta} &= -\sin(\theta) \notag\\
\frac{\partial F_2}{\partial \lambda} &= - LC \left(\lambda +\frac{1}{2}\left(\frac{G}{C}+\frac{R}{L}\right)\right) 
\label{eq:partialDerivatives}
\end{align}
 where we have used trigonometric identities on the product of two sines to simplify the expressions in $F_1$. In particular, we are interested in the influence of $G_L$ in the complex eigenvalue with the smallest natural frequency, representing the first resonance, and the relevant one for inverter-based generation, so our operating point of interest corresponds to $\theta_* = \frac{\pi}{n+1}$. Then, using the expressions in~\eqref{eq:partialDerivatives}, the relationship in~\eqref{eq:sensitivities} between the variations in the eigenvalues $\Delta \lambda$ and the changes in the load $\Delta G_L$ is computed in~\eqref{eq:finalRelationshipLambdaG}.
\begin{figure*}[h]
\begin{align}
\frac{\Delta \lambda}{\Delta G_L} &= \frac{L\left(\lambda_*+\frac{R}{L}\right)\frac{1}{2}}{LC \left(\lambda_* +\frac{1}{2}\left(\frac{G}{C}+\frac{R}{L}\right)\right)}\cdot
\frac{\cos((n-j)\theta_*)+1}{\frac{-(n-j)\cos(\theta_*)\sin((n-j)\theta_*)-n\sin(n \theta_*)}{\sin(\theta_*)}-\cos((n-j)\theta_*)-1} \notag\\
&\approx
\frac{1}{2C} \cdot\frac{\cos((n-j)\theta_*)+1}{\frac{-(n-j)\cos(\theta_*)\sin((n-j)\theta_*)-n\sin(n \theta_*)}{\sin(\theta_*)}-\cos((n-j)\theta_*)-1}
\label{eq:finalRelationshipLambdaG}
\end{align}
\end{figure*}
Given the large magnitude of the imaginary part of the complex eigenvalues and the comparatively low values of $R$ and $G$, the first term can be approximated by $1/2C$. Notice how the sensitivity barely depends on the values of $L$ and $R$. Moreover, the optimal value of $j$ (that is, the location of the load), does not depend on the value of $C$. In fact, the second term in the expression only depends on $n$ and $j$, and can clearly be evaluated independently of the line parameters. It can be easily seen that, for $j \leq n$, the relationship is mostly real, and thus a change in the load leads to a change in just the real part of the eigenvalues, with the imaginary component remaining unaffected. Moreover, this sensitivity is always negative, given that the numerator is always positive and the terms in the denominator are all non-positive. This implies that the real part of the first complex eigenvalue becomes more negative as $G_L$ increases, and therefore this resonance is more damped, regardless of the location of the load. The denominator is never $0$, so the expression is well defined.  

Furthermore, it can be shown by means of lengthy computations that the derivative of the right hand side of~\eqref{eq:finalRelationshipLambdaG} over $j$ is always negative for $j \leq n$, that is, the expression is monotonically decreasing as a function of $j$, which implies that the closer the load is to the middle point\footnote{With our definition of $j$, the case of $j=n$ corresponds to $z = \frac{n+1}{2}$, which is the center of the line for odd values of $n$. For even values, the exact center point does not correspond to a physical node, but nonetheless $\lceil \frac{n+1}{2} \rceil$ and $\lfloor \frac{n+1}{2} \rfloor$ are the closest to the center point and the optimal locations for the load.}, the larger the damping is. The derivative can be shown to be 0 at $n=j$, and is positive for $j>n$, as expected from the symmetry in the $A$ matrix. Hence, the load at the center point maximizes the sensitivity of the magnitude of $\Delta \lambda / \Delta G_L$ and therefore maximizes the damping of the first resonance.

\begin{rem}
\label{remarkLoadMiddle}
For the case of $j=n$, it can be seen that $\theta = 2\pi/(n+1)k$, for $k=1,2,..$ satisfies~\eqref{eq:charPol_overJ} regardless of the value of $G_L$, that is, the even resonances of the original system (i.e., those with even values for $k$ in the definition of $\theta$ in~\eqref{lambdaComplex}) are not modified by the presence of the load. This already hints a limitation of placing a load in a single location: it will not get rid of certain resonances no matter how large the load is. Intuitively speaking, distributing loads along the line will make all the eigenvalues real, by making the $A$ matrix diagonally dominant. This is however, from a practical perspective, much more complicated to implement that the connection of a single load along the line.
\end{rem}
\begin{rem}
\label{remarkLargeG}
The previous analysis was limited to small load values, since it resorted to a series expansion around the operating point of the unloaded case (which are nonetheless the values of interest, given the limitations in terms of power rating). On the other spectrum, for large values of $G_L$, the previous expressions can be largely simplified, leading to the characteristic polynomial:
$$ \Delta_{2n+1} \approx \left(\lambda+\frac{G_L}{C}\right)\Delta_j\Delta_{2n-j} $$
Many interesting remarks can be drawn from this simple expression. First, besides the real eigenvalue $\lambda=-G_L/C$, the eigenvalues satisfy $\Delta_j=0$ and/or $\Delta_{2n-j}=0$. This implies that, regardless of the load value and its location, resonances of different frequencies will still exist. Hence, as pointed out in the previous remark, placing large loads in some 
particular nodes does not eliminate all resonances, no matter how large they are (see the example in the next section and the associated root locus for further clarification); instead, it seems to be more beneficial to distribute loads along the grid.
\end{rem}
\begin{rem}
As an academic exercise, similar derivations could be carried out for the other complex eigenvalues beyond the first resonance, in order to identify the optimal location to maximize the damping of each resonance. As we will see in the example section, we can conjecture it follows a clear pattern. 
\end{rem}
\section{An academic example}
To illustrate our results, we consider a \SI{100}{km} transmission line with R = $0.02$ \si{\Omega / km}, L = $0.5 \cdot 10^{-3}$ \si{H/km}, C = $0.4 \cdot 10^{-6}$ \si{F/km} and G = $0$ \si{S/km}. These values refer to a typical \SI{110}{kV} transmission line. To highlight the versatility of our method, the line is divided in $n=60$ $\pi$-sections of equal length. Therefore, the system has $121$ states variables.

We first compute numerically the eigenvalues once a load of $G_L = 1/100$ is located at each node along the line, and compare it against the unloaded line. For an eigenvalue of the form $\lambda=a+jb$, we define the damping factor as usual, that is:
$$ \sigma = \frac{-a}{\sqrt{a^2+b^2}}$$

\begin{figure}[h]
\includegraphics[width=1\columnwidth]{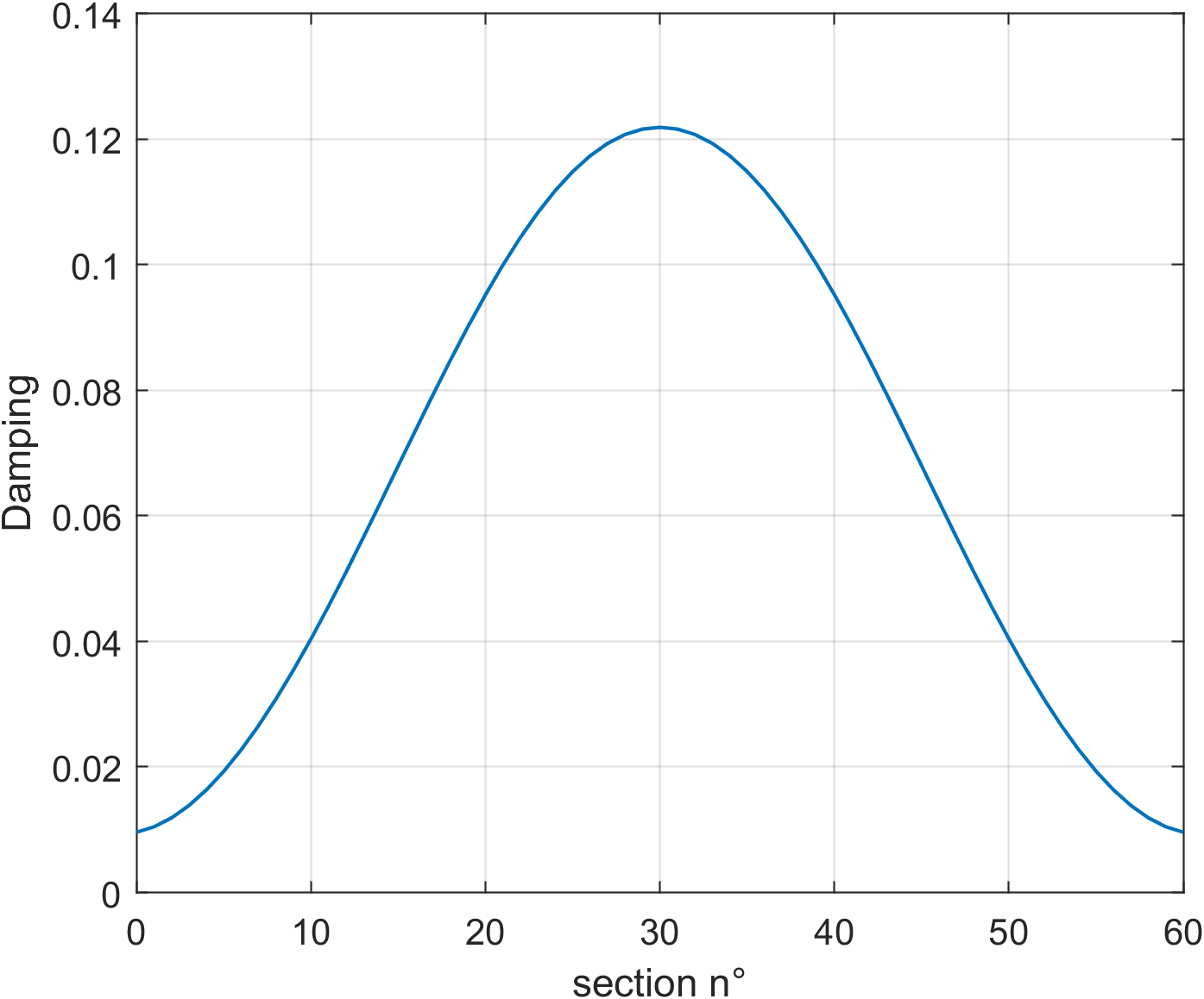}
\caption{Damping factor of the first resonance as the load is connected along  the transmission line.}
\label{fig:damping1stResonance}
\end{figure}

Figure~\ref{fig:damping1stResonance} shows the evolution of the damping factor of the complex eigenvalue with the lowest resonant frequency when the load is located at different points. As expected from the theoretical results in Section~\ref{sec:loadPlacement}, it can be clearly observed not only how the central location maximizes the damping factor, but also that as the load nears the center point, the damping factor increases. This translates into a clear guideline for grid restoration: the loads to be connected should be as close as possible to the center point of the transmission line to be energized.

\begin{figure}[h]
\includegraphics[width=1\linewidth]{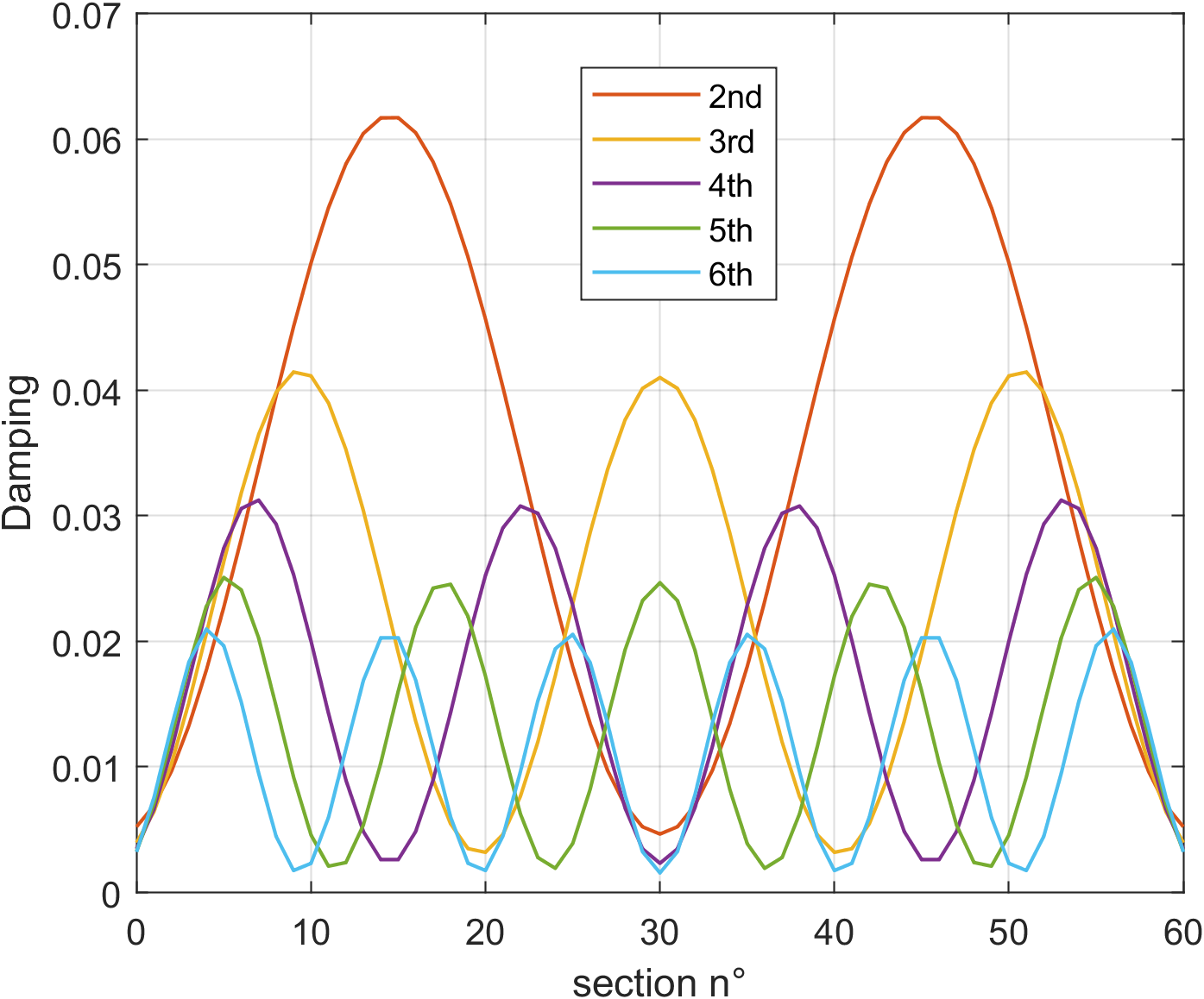}
\caption{Damping factor of the higher resonances as the load is connected along  the transmission line.}
\label{fig:dampingHigherResonances}
\end{figure}

\begin{figure}[h]
\includegraphics[width=1\linewidth]{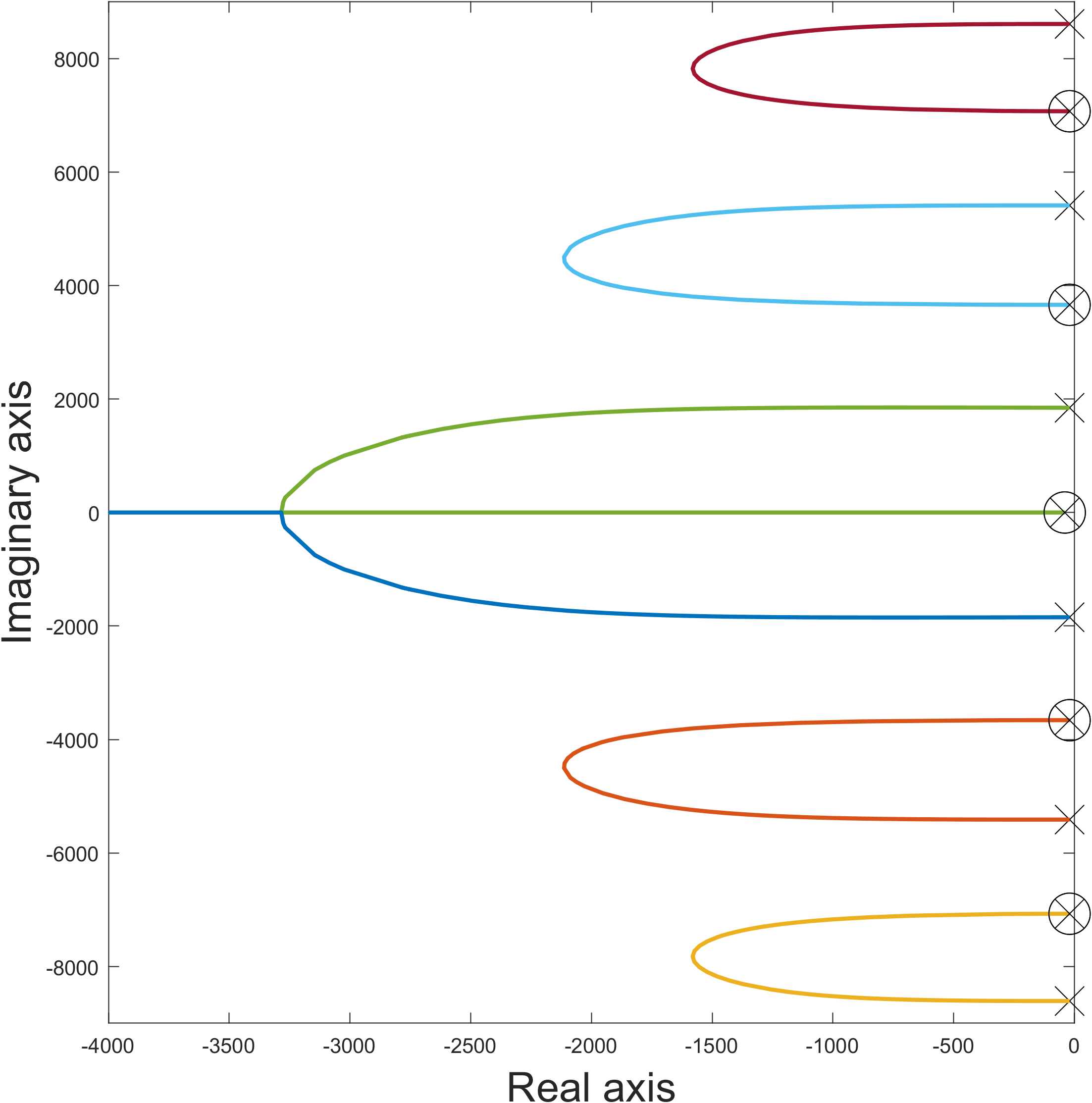}
\caption{Root locus with a load located at the center point.}
\label{fig:rootLocus}
\end{figure}

We also compute the evolution of the damping factors for the higher resonances, displayed in Figure~\ref{fig:dampingHigherResonances}. Interestingly, while the optimal location for the first resonance was at the section $z=n/2$, we observe anecdotally for this example that the optimal location to maximize the damping ratio of the second resonance is at $z=n/4$, the third resonance is maximally damped when the load is at $z=n/6$, the fourth at $z=n/8$, and so on. The corresponding analytical analysis has not been carried out since these questions are in general less relevant from a practical perspective, so no claims can be made on the generalization of this insight. The plot also shows how the damping of the even resonances (2\textsuperscript{nd}, 4\textsuperscript{th}, etc) barely changes when the load is placed in the middle, as mentioned before in Remark~\ref{remarkLoadMiddle}. Moreover, the load at the beginning or end of the line seems to be the worst location, since it barely improves the damping of any resonance. 
Finally, we display in Figure~\ref{fig:rootLocus} the root locus of the system when the load is placed in the middle point of the line (with 'o' denoting the zeros and 'x' denoting the open-loop poles, as usual), zooming into the first resonances. As concluded by our study, a load that is large enough would make the first complex eigenvalue purely real, but it would not influence at all the location of the even resonances. Interestingly, we can see how the other odd resonances end up at the location of the even resonances, which could be predicted from the expression in Remark~\ref{remarkLargeG}. 






\section{Conclusions and Outlook}
This paper has explored the effect of placing a load in a transmission line to improve damping during the energization of transmission lines. We have mathematically proven that the optimal location happens to be exactly in the center point. The results provide clear rules that are easy to implement for a system operator.
Even though this paper focused on the simple case of a transmission line, it is nonetheless a realistic starting point for certain system operators. Given that grid restoration strategies are moving towards backbone structures instead of cell-based, it is expected that more complex grids are energized in a single shot. Future work will thus leverage the framework used here to consider the effect of more complex topologies on the damping of existing resonances, given that the initial topology of the network can typically be selected by the system operator. Likewise, it is necessary to understand how the optimal location of the load depends on the properties of the network graph.

\nocite{dorfler2018electrical,yueh2008explicit,alvarez2005some}

\bibliography{root_bib}

\end{document}